\documentclass[prd,aps,amsmath,twocolumn,superscriptaddress,amssymb,reprint,nofootinbib]{revtex4-2}
\pdfoutput=1
\usepackage{graphicx,array}
\usepackage{hyperref}
\usepackage{color}
\usepackage{amsmath,amssymb,slashed,latexsym}
\usepackage{bm}
\usepackage{braket}
\usepackage{placeins}
\usepackage{adjustbox}
\usepackage{subcaption}
\usepackage{enumitem}

\def\beq{\begin{equation}}
\def\eeq{\end{equation}}
\def\bea{\begin{eqnarray}}
\def\eea{\end{eqnarray}}

\usepackage{array}
\newcolumntype{P}[1]{>{\centering\arraybackslash}p{#1}}
\newcolumntype{M}[1]{>{\centering\arraybackslash}m{#1}}

\definecolor{darkblue}{cmyk}{1,0.4,0,0.3}
\definecolor{violet}{cmyk}{0,1,0,0.2}
\hypersetup{colorlinks, bookmarksnumbered, citecolor=darkblue, linkcolor=darkblue, pdfstartview=FitH, urlcolor=darkblue, linktocpage}

\begin{document}

\title{Scrutinizing the Primordial Black Holes Interpretation of PTA Gravitational Waves and JWST Early Galaxies}

\author{\large Yann Gouttenoire}
\email{yann.gouttenoire@gmail.com}
\affiliation{ School of Physics and Astronomy, Tel-Aviv University, Tel-Aviv 69978}

\author{\large Sokratis Trifinopoulos}\email{trifinos@mit.edu}
\affiliation{Center  for  Theoretical  Physics,  Massachusetts  Institute  of  Technology,  Cambridge,  MA  02139,  USA}

\author{\large Georgios Valogiannis}\email{gvalogiannis@g.harvard.edu}
 \affiliation{
 Department of Physics, Harvard University, Cambridge, MA, 02138, USA\\
}

\author{\large Miguel Vanvlasselaer}\email{miguel.vanvlasselaer@vub.be}
\affiliation{Theoretische Natuurkunde and IIHE/ELEM, Vrije Universiteit Brussel, \& The  International Solvay Institutes, Pleinlaan 2, B-1050 Brussels, Belgium}

\begin{abstract}

Recent observations have granted to us two unique insights into the early universe: the presence of a low-frequency stochastic gravitational wave background detected by the NANOGrav and Pulsar Timing Array (PTA) experiments and the emergence of unusually massive galaxy candidates at high redshifts reported by the James Webb Space Telescope (JWST). In this letter, we consider the possibility that both observations have a common origin, namely primordial black holes (PBHs) in the mass range between $10^{6}~M_{\odot}$ and $10^{13}~M_{\odot}$. While superheavy PBHs act as seeds for accelerated galaxy formation capable of explaining the JWST extreme galaxies, they can also form binary mergers that source gravitational waves which can be potentially identified as the PTA signal. The analysis is performed taking into account the constraints on the relevant region of the PBH parameter space including the novel bound imposed by the Ultraviolet Luminosity Function of galaxies observed by the Hubble Space Telescope. We conclude that PTA's and JWST's interpretations in terms of PBH binary mergers and Poissonian gas of PBHs, respectively, are strongly excluded.

\end{abstract}

\maketitle

\section{Introduction}
\label{sec:intro}

The NANOGrav collaboration \cite{NANOGrav:2020bcs,NANOGrav:2023gor} combined with the other Pulsar Timing Array (PTA) Collaborations~, EPTA~\cite{Chen:2021rqp,Antoniadis:2023rey}, PPTA~\cite{,Goncharov:2021oub,Reardon:2023gzh}, CPTA~\cite{Xu:2023wog} and IPTA~\cite{Antoniadis:2022pcn} have recently released further evidence for the Hellings-Downs angular correlation in the common-spectrum process. This points toward the existence of a Gravitational Wave (GW) background in the nHz range permeating the universe, sparking a lot of new physics speculation~\cite{NANOGrav:2023hvm,Antoniadis:2023zhi}.

A different lens through which we can gain glimpses into the early cosmic evolution is presented by the James Webb Space Telescope (JWST). Initial observations have reported photometric evidence of massive galaxies at unexpectedly high redshifts $7<z<12$~\cite{Adams2022,Naidu2022,Finkelstein2023,2023Natur.616..266L}. A subset of them has been recently spectroscopically confirmed~\cite{curtislake2023spectroscopic,Robertson:2022gdk} and large cosmological hydrodynamical simulation demonstrated compatibility with existing models of galaxy formation~\cite{Keller:2022mnb,McCaffrey:2023qem}. However, the status of extreme galaxy candidates with stellar mass as high as $10^{11}~M_{\odot}$~\cite{2023Natur.616..266L} still remains under investigation and if their distance estimates prove accurate, they would pose a significant challenge to $\Lambda$CDM itself~\cite{Boylan-Kolchin:2022kae,Lovell:2022bhx,Haslbauer:2022vnq}. Also in this case, explanations beyond the standard cosmological paradigm have been postulated~\cite{Liu:2022bvr,Gong:2022qjx,Hutsi:2022fzw,Menci:2022wia,Biagetti:2022ode,Yuan:2023bvh,Ilie:2023zfv,Parashari:2023cui,Jiao:2023wcn,Guo:2023hyp,Su:2023jno}.

Primordial black holes (PBHs) emerge as one of the most long-studied scenarios~\cite{Hawking:1971ei,Carr:1974nx,Meszaros:1974tb,Carr:1975qj,Chapline:1975ojl,Carr:2016drx,Green:2020jor}, capable of leaving distinctive imprints on cosmic history. Depending on the fraction of their abundance with respect to the total dark matter (DM) $f_{\rm PBH} = \Omega_{\rm PBH}/\Omega_{\rm DM}$, the spectrum of possible PBH masses $M_{\rm PBH}$ spans an enormous range which has been tested by various experiments (for comprehensive reviews see refs. \cite{Sasaki:2018dmp,Carr:2020gox}). PBHs are expected to form binaries~\cite{Quinlan:1987qj,Mouri:2002mc,Bird:2016dcv,Sasaki:2016jop,Kashlinsky:2016sdv,Ali-Haimoud:2016mbv,Raidal:2017mfl,Raidal:2018bbj,Sasaki:2018dmp}, which once decoupled from the universe evolution, continuously emit GWs until a last spectacular burst when they finally merge. Mergers of stellar mass BHs have been observed by terrestrial interferometers~\cite{LIGOScientific:2016dsl, LIGOScientific:2020ibl}, with the potential of a primordial origin not ruled out~\cite{Sasaki:2016jop,Hall:2020daa,Jedamzik:2020ypm,DeLuca:2020qqa,Franciolini:2021tla, Romero-Rodriguez:2021aws}.

PBHs heavier than $100~M_{\odot}$ are of particular interest due to their influence on the growth and formation of structures. For example, it is a well-known fact that supermassive black holes (SMBHs) with masses above $10^{5}~M_{\odot}$ reside within galactic nuclei~\cite{Ferrarese:2004qr,Gultekin:2009qn,Kormendy:2013dxa}. Naturally, it has been proposed that PBHs can be their progenitors reaching these masses either by merging and accretion~\cite{Bean:2002kx,Kawasaki:2012kn,Ali-Haimoud:2016mbv,Clesse:2016vqa,Clesse:2017bsw,Serpico:2020ehh}  or direct collapse of primordial fluctuations~\cite{Nakama:2016kfq,Nakama:2017xvq}. In the latter case, supermassive PBHs are constrained to be less than $\mathcal{O}(0.1 \%)$ of DM and since they would already be present from the dawn of the matter-domination, they can act as cosmic seeds boosting galaxy formation~\cite{Carr:2018rid,Inman:2019wvr}. Moreover, a subdominant component of DM can consist of stupendously large PBHs, which are heavier than $10^{12}~M_{\odot}$~\cite{Carr:2020erq} and may traverse the intergalactic medium.

In this letter, we focus on supermassive PBHs in the mass range $10^{6}<M_{\rm PBH}/M_{\odot}<10^{13}$ and assess the viability of the following two scenarios: i) they bind in binary systems which leads to late-time merging and radiation of GWs that are detectable by PTA experiments~\cite{Atal:2020yic,Depta:2023qst},
 ii) they are responsible for the generation of the early massive galaxies reported by JWST~\cite{Liu:2022bvr,Hutsi:2022fzw}.
 For the first case, we calculate the GW energy spectrum induced by binary mergers and then perform a Bayesian analysis to determine the posterior distribution compatible with the NANOGrav 15-year \cite{NANOGrav:2020bcs} and IPTA DR2 \cite{Antoniadis:2022pcn} signals. 
 For the second case, the accelerated galaxy formation in the presence of PBHs is investigated using both the Poisson and seed effects~\cite{Carr:2018rid} and we identify the PBH populations that can sufficiently seed a large number of massive galaxies at $z \sim 10$. 

The results of our analysis are presented combined with all relevant cosmological and astrophysical constraints. Notably, we consider for the first time in the PBH literature a constraint inferred by measurements of the Hubble Space Telescope (HST) and encoded in the so-called UV galaxy luminosity function (UV LF)~\cite{2018ApJ...855..105O,2021AJ....162...47B}. Observations of UV-bright galaxies collected by the Hubble Space Telescope (HST)~\cite{Bouwens:2014fua,2015ApJ...810...71F,Atek:2015axa,Livermore:2016mbs,2017ApJ...843..129B,2017ApJ...838...29M,2018ApJ...854...73I,2018ApJ...855..105O,Atek:2018nsc,2020ApJ...891..146R,2021AJ....162...47B} trace the universe at redshifts $z=4-10$, significantly overlapping with the ones performed by the JWST. In the context of our combined analysis, we find that the HST UV LF together with the large-structure constraints (LSS) rule out the PTA's interpretation in terms of GW from PBH mergers as well as the JWST's interpretation in terms of Poissonian density fluctuation sourced by PBHs.

\section{Fitting the PTA signals with gravitational waves from PBH binary mergers}
\label{sec:GW_mergers}

Supermassive PBHs can form in the early universe at redshift $z_{\rm f}$ with a mass $M_{\rm PBH} \simeq \big(5 \times 10^{12}/(1+z_{\rm f})\big)^2M_{\odot}$~\cite{Sasaki:2018dmp}. Immediately after their formation, PBHs are sparsely distributed in space, with mean separation much larger than the Hubble scale, $l_{\rm PBH}(z_{\rm f}) \gg H^{-1}(z_{\rm f})$. However the mean separation between PBHs, $l_{\rm PBH}(t) \propto t^{1/2}$, falls below the Hubble distance $H^{-1} \propto t$ before matter-radiation equality and several PBHs can coexist in the same Hubble patch, leading to the possibility of creating pairs of PBHs. In principle, PBHs binaries can form both before or after matter-radiation equality (early~\cite{Nakamura:1997sm} and late-time~\cite{Quinlan:1987qj,Mouri:2002mc,Bird:2016dcv} formation), however the early contribution dominates for black holes of primordial origins, e.g. \cite{Raidal:2018bbj,Sasaki:2018dmp}. We present the calculation of the distribution of binaries and of the associated merging rate $\mathcal{R}(z)$, in Appendix \ref{app:bin_for}.\footnote{For the possibility of initial clustering of PBHs and the implication for the PTA signal see ref.~\cite{Depta:2023qst}.}
After the creation of the binary system, the two PBHs will orbit around each other and merge approximately within a time \cite{Peters:1963ux} 
\begin{align} 
t_{\rm m} 
&\simeq \frac{3}{170}\frac{1}{G^3 M^3_{\rm PBH}} a^4 (1-e^2)^{7/2}~,
\end{align}
which depends on the eccentricity  $e$ of the orbit of the two PBHs and its major axis  $a$. $G$ is the gravitation constant. Circular binaries with abundance $f_{\rm PBH} \simeq (10^{13}M_{\odot}/M_{\rm PBH})^{5/16} $ will typically merge today $t_{\rm m} \simeq 13.8$ Gyr. 
\paragraph*{\textbf{Gravitational waves from PBH binary mergers.}}
The energy density of the stochastic gravitational waves from PBH binaries reads  \cite{LIGOScientific:2016fpe,PhysRevLett.116.131102}
\begin{equation}
h^2 \Omega_{\rm GW} = \frac{f}{\rho_c/h^2 } \int_0^\infty dz \frac{\mathcal{R}(z)}{(1+z)H(z)} \frac{dE_{\rm GW}(f')}{df'} \bigg|_{f'=(1+z)f}~,
\label{eq:GW_signal}
\end{equation}
where $\rho_c/h^2 = (3.0~\rm meV)^4$ is the Hubble-rescaled critical density at present time, $dE_{\rm GW}(f')/df'$ is the GW power emitted by a circular and GW-driven binary, $f'$ is the rest frame frequency and $f=f'/(1+z)$ the observed one. 
We find a fit for the GW signal with the following power broken-law
\begin{align} 
 \label{eq:fit}
\Omega_{\rm GW} h^2 \simeq    \, \Omega_{\rm peak} S(f) \Theta(2f_{\rm peak} - f)~,
\end{align} 
where the spectral function is
\begin{align} 
 S(f) =   \,\frac{f_{\rm peak}^b f^a }{\bigg(b f^{\frac{a+b}{c}}+ a f_{\rm peak}^{\frac{a+b}{c}}\bigg)^c}~,
 \end{align} 
the peak amplitude is
\begin{align}
 \Omega_{\rm peak}\simeq   \, 0.05 f_{\rm PBH}^3 \bigg(1.5\frac{M_{\rm PBH}}{10^{12} M_{\odot}}\bigg)^{-0.3}~,
\end{align}
and the peak frequency is $f_{\rm peak } \simeq  \, 5000 M_{\odot}/M_{\rm PBH}$. The fitting parameters are $a = 0.7, \quad b = 1.5, \quad c = 0.9$. We refer the reader to Appendix \ref{app:GW_PBH}  for more details. 
The step function $\Theta$ is introduced to cut the GW spectrum at frequencies above the ring-down frequency $f > 2f_{\rm peak}$.
On top of this estimate, we consider the following two effects:
 \paragraph{Environmental effects.}
    At very low frequencies (IR), the assumption of GW-driven energy loss breaks down and interactions with the environment generate additional energy losses which change the spectral shape of the GW background \cite{Kelley:2017lek,Burke-Spolaor:2018bvk}. Following \cite{Rosado:2015epa}, we discard the region of frequencies lower than 
\bea 
\label{eq:f_min}
f_{\rm min} = \bigg(\frac{T_{\rm max}}{\delta_2}\bigg)^{\!-\tfrac{3}{8}} \hspace{-0.2cm}~~, \qquad \delta_2 = \frac{5}{256 \pi^{\!\frac{8}{3}} (G M_c)^{\!\frac{5}{3}}}  ~,
\eea 
where the choice $T_{\rm max} = 75$ Myr corresponds to the transition from the stellar-scattering dominated phase to the GW dominated one \cite{Rosado:2015epa}. \\
 \paragraph{Discrete versus continuous signal.}
 Eq.~\eqref{eq:GW_signal} assumes a smooth distribution of sources. However, above a frequency $f_{\rm max}$, the number of sources per frequency bins $N(f, \Delta f)$ (we explain in Appendix \ref{app:GW_PBH} how to compute the function $N$) can become smaller than one $N(f, \Delta f)<1$ and the assumption of continuity in Eq.~\eqref{eq:GW_signal} breaks down \cite{Rosado:2015epa}. The stochastic signal is replaced by a pop corn noise which could in principle be subtracted~\cite{Taylor:2020zpk}. We follow the approach of ref.~\cite{Sesana:2008mz} to determine when $N(f, \Delta f)<1$.

\section{PBH-enhanced Structure formation and the current JWST observations}
\label{sec:structure_formation}

Assuming a monochromatic mass function, the PBH population can be parameterized by the common mass, $M_{\rm PBH}$, and the DM fraction $f_{\rm PBH}$. At scales where the discrete nature of PBHs becomes relevant, there are two different effects that can influence structure formation.

\paragraph*{ \textbf{Poisson effect.}} In a region of mass $\tilde{M}$ the condition $f_{\rm PBH}>M_{\rm PBH}/\tilde{M}$ implies that one expects to find more than one PBH $\tilde{N}_{\rm PBH}>1$. For initially Poisson-distributed PBHs the fluctuations in their number $\sqrt{\tilde{N}_{\rm PBH}}$ generate isocurvature density perturbations of magnitude $\delta_{\rm{PBH},i} \approx (f_{\rm PBH} M_{\rm PBH}/\tilde{M})^{1/2}$ (see appendix A in ref.~\cite{Papanikolaou:2020qtd} for a detailed derivation).
The perturbations evolve linearly right after matter-radiation equality~\cite{Carr:1974nx} (i.e. $z < z_{\rm eq} \approx 3400$). As a result, the matter power spectrum is modified at smaller scales by the addition of an isocurvature component~\cite{Inman:2019wvr}
\begin{align} \label{eq:power_spectrum}
    &P(k) = P_{\rm ad}(k) + P_{\rm iso}(k)~, \notag \\
    & P_{\rm iso}(k) \simeq  \begin{cases}
    \frac{(f_{\rm PBH} D(0))^2}{\bar{n}_{\rm PBH}}~,& \text{if } k\leq k_{\rm cut} \\
    0~,              & \text{otherwise}
\end{cases}~,
\end{align}
\noindent
where $P_{\rm ad}(k)$ is the adiabatic mode in $\Lambda$CDM\footnote{We generate a prediction for the linear matter power spectrum using the Boltzmann solver \textsc{CAMB} \cite{Lewis:1999bs}, for a fiducial $\Lambda$CDM cosmology corresponding to the following {\it Planck} 2018 \cite{Planck:2018vyg} values : $n_s=0.9649$, $\sigma_8=0.8111$, $\Omega_b=0.0493$, $\Omega_{\rm DM}=0.266$ and $h=0.6736$.}, $\bar{n}_{\rm PBH} = f_{\rm PBH} \rho_{\rm crit,0} \Omega_{\rm DM}/M_{\rm PBH}$ is the co-moving average number density of PBHs, and $D(z) \simeq \left(1+\frac{3\gamma}{2\alpha_-} \frac{1+z_{\rm eq}}{1+z}\right)^{a_-}$ is the growth factor with $a_- = \left(\sqrt{1+24\gamma}-1\right)/4$ and $\gamma = (\Omega_{\rm DM} - \Omega_{\rm b})/(\Omega_{\rm DM} + \Omega_{\rm b})$. 
The isocurvature term is truncated at the scale $k_{\rm cut}$ which is the scale where we expect the linear Press-Schechter theory (PS)~\cite{1974ApJ...187..425P} to break down. Two different cut-off scales $k_{\rm cut}$ have been discussed, i.e. the inverse mean separation between PBHs, $\bar{k}_{\rm PBH} = (2 \pi^2 \bar{n}_{\rm PBH})^{1/3}$~\cite{Inman:2019wvr,DeLuca:2020jug,Hutsi:2022fzw},
 and the approximate scale $k_{\rm NL} \approx (\bar{n}_{\rm PBH}/f_{\rm PBH})^{1/3}$~\cite{Liu:2022bvr}, where non-linear dynamics (see the seed effect below and the case of mode mixing in ref.~\cite{Liu:2022okz}) starts to dominate. In the absence of a reliable description of the transition between the linear and the non-linear regimes, we perform our analysis reporting both benchmarks, the conservative scenario $k_{\rm cut}=\bar{k}_{\rm PBH}$ being shown in fig.~\ref{fig:PBH_exclusion} and the more aggressive one being reported to fig. 2 in the Appendix \ref{app:kPBH_cut}.

Equipped with the enhanced power spectrum, we utilize the Sheth-Tormen (ST) ~\cite{Sheth:1999mn,Sheth:1999su} modification to the PS formalism with a top-hat window function (see Appendix \ref{app:HMF}) and calculate the halo mass function $dn(M_h,z)/dM_h$. The expected number density of galaxies with stellar mass above the observational threshold, $M_{\star}^{\rm obs}$, is given by~\cite{Boylan-Kolchin:2022kae}
\begin{equation}\label{eq:number_density}
    n_{\rm gal}(M_\star \geq M_{\star}^{\rm obs}) = \int_{M^{\rm cut}_h}^{\infty} \frac{dn(z_{\rm obs}, M_h)}{dM_h} dM_h~.
\end{equation}
\noindent
Under the assumption that each dark-matter halo contains a single central galaxy, the relation between the halo and total stellar mass is $M_h (M_\ast) = M_\ast / (f_{\rm b} \epsilon_\ast)$, where $f_b = \Omega_{\rm b}/(\Omega_{\rm DM}+\Omega_{\rm b}) = 0.157$ is the baryon fraction and $\epsilon_\ast$ the star formation efficiency. The lower limit of the integral in eq.~\eqref{eq:number_density} is defined then as ${M^{\rm cut}_h} = M_h (M_{\star}^{\rm obs})$. 
The JWST signature can be expressed as $ n_{\rm gal}(M_\star \geq 10^{10.8}~M_{\odot}) \simeq 10^{-5} \rm Mpc^{-3}$ at $z_{\rm obs} \sim 8$~\cite{2023Natur.616..266L}.

\paragraph*{ \textbf{Seed effect.}} In the opposite limit, $f_{\rm PBH}<M_{\rm PBH}/\tilde{M}$, the PBHs make up only a small fraction of dark matter and form isolated halos. The density perturbations in this case are $\delta_{\rm{PBH},i} \approx M_{\rm PBH}/\tilde{M}$,  binding gravitationally regions of mass $\tilde{M}(M_{\rm PBH},z) \simeq z_{\rm eq} M_{\rm PBH}/(z+1)$.
Due to its highly non-linear nature, this effect can be examined properly only using simulations~\cite{Inman:2019wvr}. 
However, following the calculation of refs.~\cite{Liu:2022bvr,Su:2023jno} we can determine the part of the parameter space that can potentially accommodate accelerated early galaxy formation compatible with the JWST observations solely with the seed effect. The three basic requirements are: (i) $f_{\rm PBH}<M_{\rm PBH}/\tilde{M} = (z+1)/z_{\rm eq}$, (ii) $\bar{n}_{\rm PBH} > 10^{-5} \rm Mpc^{-3}$, and (iii) $ \tilde{M}(M_{\rm PBH},z_{\rm obs}\sim 8) \simeq M_h(M_{\ast}^{\rm obs}\sim 10^{11}M_{\odot})$, assuming also in this case that the whole bound region hosts a single galaxy. 

\section{Observational constraints}\label{sec:constraints}

In this section we review the constraints on the PBH populations with masses above $10^6~M_{\odot}$ and introduce the UV LF bound on the PBH parameter space.

\begin{enumerate}

    \item \textbf{Large-scale structure and dynamical friction.} As discussed in Section \ref{sec:structure_formation}, PBHs can significantly contribute to the formation of cosmic structures. In fact, the non-observation of different types of structures can be used to constrain population of PBHs~\cite{Carr:2018rid}.
    We comment however that those constraints are approximative as they primarily apply to the bulk of various types of structures. For specific cases (e.g. the JWST early galaxies), the limits may vary within an order of magnitude.

    The dynamical friction (DF) limit concerns the accumulation of halo BHs into the galactic nucleus. The effect is induced due to dynamical friction by stellar populations and the subsequent merging of those BHs in the nucleus would result into SMBHs heavier than the currently observed ones unless $f_{\rm PBH}$ is rather small. The relevant derivation can be found in ref.~\cite{Carr:1997cn}.
    
    \item \textbf{CMB $\mu$ distortion.}
    
    The PBH formation from large-amplitude primordial fluctuations leaves imprints in the CMB spectrum, called $\mu$ distortion~\cite{Chluba:2012we}. The defining observable $\mu$ is strictly constrained by the COBE/FIRAS measurements to be smaller than $4.7\times 10^{-5}$~\cite{Fixsen:1996nj,Bianchini:2022dqh}, which translates into a strong bound on the PBH abundance over the mass range $10^5~M_{\odot} < M_{\rm PBH} < 10^{12}$. If the primordial fluctuations are Gaussian then practically all PBH models of interest are excluded~\cite{Nakama:2017xvq}.~\footnote{See however the models of refs.~\cite{Hawking:1987bn,Garriga:1992nm,Garriga:2015fdk,Deng:2016vzb,Deng:2017uwc,Deng:2018cxb,Nakama:2016kfq,Kawasaki:2019iis,Cotner:2019ykd,Kitajima:2020kig,Kawana:2021tde,Kasai:2022vhq,Chang:2018bgx,Lu:2022jnp,Domenech:2023afs}, which postulate different PBH formation mechanisms and may avoid bounds from CMB spectral distortion altogether.} On the other hand, if one allows for significant non-Gaussianities (NG) in the curvature power spectrum the bound can be avoided. In this work, we adopt the phenomenological description of ref.~\cite{Nakama:2016kfq} and characterize NGs by the parameter $p \leq 2$ (where $p=2$ corresponds to the Gaussian case). All the relevant expressions can be found in Appendix \ref{app:mu-distortion}.
    
    \item \textbf{HST UV luminosity function.}   
    
    The UV LF of galaxies observed by the Hubble Space Telescope (HST)~\cite{Bouwens:2014fua,2015ApJ...810...71F,Atek:2015axa,Livermore:2016mbs,2017ApJ...843..129B,2017ApJ...838...29M,2018ApJ...854...73I,2018ApJ...855..105O,Atek:2018nsc,2020ApJ...891..146R,2021AJ....162...47B} has emerged as a sensitive early universe probe to constrain not only $\Lambda$CDM~\cite{Bouwens:2014fua,2016MNRAS.460..417S,Naidu:2019gvi,Mason:2015cna,Sahlen:2021bqt,Sabti:2021unj,Sabti:2021xvh}, but also a broad portfolio of extensions about it \cite{Menci:2020ybl,Bozek:2014uqa,Schultz:2014eia,Dayal:2014nva,Corasaniti:2016epp,Menci:2017nsr,Menci:2018lis,Rudakovskyi:2021jyf,Yoshiura:2020soa,Chevallard:2014sxa,Sabti:2020ser}.     
    As it was recently pointed out~\cite{Sabti:2023xwo}, more importantly, any models that attempt to modify the high-redshift halo mass function in relation to the JWST excess of massive galaxies~\cite{2023Natur.616..266L} are already strongly limited by the HST UV LF, since the latter traces galaxies of the same redshifts and distances as the ones in the JWST sample. Such is the case of the massive PBHs that we consider in this work, which predict an enhancement of dark matter clustering as discussed in section \ref{sec:structure_formation}, and which have indeed been found theoretically capable of producing a desired excess of massive galaxies in the range $z_{\rm obs}=7-10$~\cite{Liu:2022bvr,Hutsi:2022fzw}. 
    
    In this letter, we take advantage of the above connection and use HST UV LF measurements \cite{2018ApJ...855..105O,2021AJ....162...47B} to constrain the properties of PBH populations using this probe. In particular, we proceed as follows: combining eqs.~\eqref{eq:power_spectrum}, \eqref{eq:number_density} and the ST formalism further explained in Appendix \ref{app:HMF}, we produce theoretical predictions for the galaxy number density above a stellar mass cut-off $M_\ast$ in the presence of a PBH population with mass $M_{\rm PBH}$ and abundance $f_{\rm PBH}$, for $\epsilon_{\star}=0.3$ at $z_{\rm obs} \sim 8$. These predictions are then contrasted against the maximum deviation from the corresponding $\Lambda$CDM result that is allowed by the HST UV LF measurements, as it was reported by the likelihood analysis of~\cite{Sabti:2023xwo} at the $95 \%$ C.L. (see fig.~3 of that work).\footnote{We note that the likelihood analysis of ref.~\cite{Sabti:2023xwo} considers model-agnostic power spectrum enhancements that were general enough to encompass modifications of the specific form in eq.~\eqref{eq:power_spectrum}, that we are using, marginalized over variations in the power spectrum amplitude and various astrophysical parameters characterizing the galaxy-halo connection.}

\end{enumerate}
 
\section{Discussion and conclusions}
\label{sec:conclusions}

\begin{figure*}[h!t]
\centering
 \includegraphics[scale=0.75]{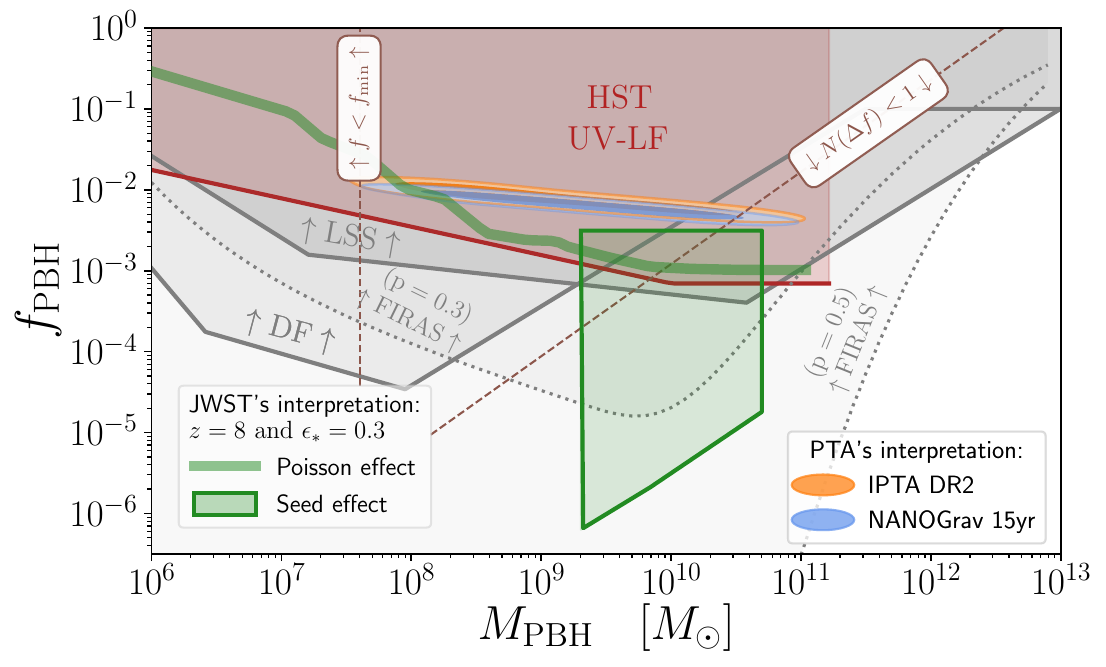}
 \caption{Illustration of the PBH parameter space in the mass region $10^{6}~M_{\odot}<M_{\rm PBH}/M_{\odot}<10^{13}~M_{\odot}$. The \textbf{green thick line} and trapeze indicate PBHs which can explain JWST's observation of early massive galaxies assuming a redshift of observation of $z\simeq 8$ and a star formation rate of $\epsilon_\ast\simeq= 0.3$. The \textbf{blue} and \textbf{orange ovals} represent the $68\%$ and $90\%$ confidence levels for PTA's interpretation in terms of GW from PBH binaries. The \textbf{two dashed brown lines} indicate where the GW-driven and smooth assumption used to derive the GW spectrum in eq.~\eqref{eq:GW_signal} break down due to environmental effects, below $f<f_{\rm min}$ defined in eq.~\eqref{eq:f_min}, and due to the number of emitting sources per frequency bin in eq.~\eqref{eq:N_Delta_f} is $N(\Delta f)<1$. The \textbf{red region} shows the constraints on PBHs due to measurements of the galaxy UV luminosity function by the Hubble Space telescope, while the \textbf{gray regions} show the constraints from large-scale structure and dynamical friction. The $\mu$-distortion bounds are depicted with the \textbf{dotted lines} for two different degrees of primordial non-Gaussianities, $p=0.3, 0.5$, and the Gaussian scenario $p=2$. The PBH isocurvature power spectrum is suppressed above $k_{\rm cut} = (2 \pi^2 \bar{n}_{\rm PBH})^{1/3}$. A different choice for $k_{\rm cut}$ is shown in fig.~\ref{fig:PBH_exclusion_app}. }
 \label{fig:PBH_exclusion}
\end{figure*}

The collective results of our analysis can be found in fig.~\ref{fig:PBH_exclusion}. This includes the regions on the $M_{\rm PBH}$-$f_{\rm PBH}$ plane that can account for both JWST and PTA signatures as discussed in sections \ref{sec:structure_formation} and \ref{sec:GW_mergers}, in conjunction with the constraints discussed in section \ref{sec:constraints}. 

We perform a Bayesian analysis using the software tool ${\tt PTArcade}$~\cite{Mitridate:2023oar} of NANOGrav 15-year \cite{NANOGrav:2023gor} and IPTA DR2 signals \cite{Antoniadis:2022pcn} in terms of GW from PBH mergers. The constraints $f_{\rm min}<f$ and $N(\Delta f)>1$ (see Appendix \ref{app:GW_PBH} for more details), signal to be stochastic, are imposed as priors and displayed as red dashed lines on Fig. \ref{fig:PBH_exclusion}. We find that PBH binaries with mass $10^8\lesssim M_{\rm PBH}/M_{\odot}\lesssim 10^{11}$ and abundance $f_{\rm PBH} \sim 10^{-2}$ can accommodate both PTA signals (68$\%$ and 95$\%$ C.L. are blue and orange ovals).

Assuming a Poissonian gas of PBHs and using linear cosmological perturbation theory and the ST approach, we model the enhancement to the galaxy abundance as a function of the grid of PBH model parameters. Two different wavelength cut-offs $k_{\rm cut}$ are used, the conservative one presented in fig.~\ref{fig:PBH_exclusion} of the main text, and the more aggressive one being reported in SuM V. The observation of massive galaxies at redshift $z\sim 8$ by the JWST is explained along the green thick line. The maximum value of $M_{\rm PBH}$ corresponds to a bound region mass of $\mathcal {O}(10^{12}M_{\odot})$, which is the typical upper theoretical limit for halo masses~\cite{Silk:1977wz,Carr:2018rid}. Additionally, using the same assumption for the linear halo evolution, we project the $95\%$ exclusion bound imposed by the UV LF on the parameter space (red shaded region). The constraints from large-scale structure and dynamical friction are also shown (gray regions).

Regarding PBHs evolving in isolation, we display within the green trapeze region of fig. \ref{fig:PBH_exclusion}, the region that could explain the JWST data with the seed effect. The upper and lower lines correspond to conditions (i) and (ii) discussed in section \ref{sec:structure_formation} (see paragraph on seed effect), respectively. The two vertical horizontal lines constrain the resulting halo masses to be $\mathcal {O}(10^{10}M_{\odot}-10^{13}M_{\odot})$, allowing for sufficient variation around $M_h(M_{\ast}^{\rm obs}\sim 10^{11}M_{\odot})$ in order to compensate for unknown uncertainties due to the simplicity of the calculation.

Finally, it is worth noticing that a common requirement for all scenarios of interest involving PBH populations in the superheavy mass region, is the formation of supermassive PBHs from NG primordial fluctuations in order to circumvent the strict bound from CMB $\mu$ distortion. In particular, in fig. \ref{fig:PBH_exclusion} we depict the $\mu$ bound given $p=0.3$ (gray dotted lines), which is marginally compatible with the depicted JWST solution based on the seed effect. This is, in fact, an extremely large degree of NGs. For instance, the benchmark $p=0.5$, which already corresponds to the infinite limit of more traditional NG measures~\cite{Nakama:2017xvq}, excludes all of the interesting PBH populations. 

\paragraph*{\textbf{Conclusions.}} In fig. \ref{fig:PBH_exclusion}, we demonstrate that the PBH populations needed to source the stochastic GW background observed in PTAs are partly excluded by LSS and decisively excluded by the UV LF constraint derived in this work. Similarly, the PBH population required to explain the JWST anomalous observation with the Poisson effect is excluded due to the same constraint. The novel UV LF bound is more rigorous than the former LSS bound of PBHs, because it has been inferred from an extended likelihood analysis of the HST data at the redshift range contemporaneous with the JWST extreme galaxy sample. The tight constraints we obtain are in agreement with the findings of ref.~\cite{Sabti:2023xwo}. 

On the other hand, the PBH solution of JWST observation based on the seed effect is in principle still viable for $f_{\rm PBH} \lesssim 10^{-3}$ up to the caveat of extreme primordial NGs. A clearer picture of the reach of the seed effect necessitates the usage of dedicated simulations in the $M_{\rm PBH}>10^7 M_{\odot}$ range, as well as an incorporating of non-linear effects in the derivation of the UV LF. Looking forward, a spectroscopic analysis will provide the final verdict on whether the JWST observations constitute a $\Lambda$CDM anomaly. Finally, a future increase in observation time of PTAs and in
number of detected pulsars might facilitate the resolution
of individual sources at larger frequencies and thus enable the more careful examination of new physics interpretations.

\begin{acknowledgments}
We would like to thank Sebastien Clesse, Edoardo Vitagliano, Daniel Eisenstein and Julian Munoz for useful discussions throughout the completion of this work, as well as Omer Katz for sparing computational resources on the TAU cluster during the preparation of this work. YG is grateful to the Azrieli
Foundation for the award of an Azrieli Fellowship. ST is supported by the Swiss National Science Foundation - project n. P500PT\_203156, and by the Center of Theoretical Physics at MIT (MIT-CTP/5538). GV recognizes partial support by the Department of Energy (DOE) Grant No. DE-SC0020223. MV is supported by the ``Excellence of Science - EOS" - be.h project n.30820817, and by the the Strategic Research Program High-Energy Physics of the Vrije Universiteit Brussel.

\end{acknowledgments}

\appendix

\section{Halo mass function}
\label{app:HMF}

We consider a spherical dark matter halo at redshift $z$, containing an average mass $\tilde{M}$ within a region of effective radius $R$. If, additionally, $\rho_{m}$ is the average matter density of the universe at $z$, it follows that
\begin{equation}\label{eq:radiusdef}
R = \left(\frac{3 \tilde{M}}{4 \pi \rho_{m}} \right)^{\frac{1}{3}}~.
\end{equation}
The root-mean square of matter density fluctuations in this region can be further calculated through a convolution of the the linear matter power spectrum, $P(k,z=0)$, with a spherical top-hat smoothing Kernel of radius $R$:
\begin{equation}\label{eq:tophat}
W\left(k R\right) = \frac{3\left[\sin(k R)-k R\cos(k R)\right]}{\left(k R\right)^3}~,
\end{equation}
which results in the (Fourier space) integral
\begin{equation}\label{eq:variance}
\sigma^2(\tilde{M}) = \int \frac{dk k^2}{2 \pi^2} W^2\left(k R\right) P(k,z=0)~.
\end{equation}

The PS theory and its generalizations \cite{1974ApJ...187..425P,1991ApJ...379..440B} postulate that the abundance of halos in the universe can be evaluated by counting all the regions of radius $R$ (and, equivalently, mass $\tilde{M}$) that have gravitationally collapsed by the time of interest $z$. According to the excursion set approach, these will be the regions with a smoothed density exceeding the critical overdensity for spherical collapse at redshift $z$, $\delta_{cr}$. For an Einstein De-Sitter (EDS) cosmology, it is always $\delta_{cr}=1.686$ \cite{1986ApJ...304...15B,1991ApJ...379..440B}, which turns out to be a very good approximation for $\Lambda$CDM cosmologies and will be adopted in this work as well. Assuming, finally, density fluctuations that are accurately described by linear perturbation theory and a Gaussian probability distribution, we can derive an analytical prediction for the comoving mean number density of halos $n_h(M)$ per logarithmic mass interval $d\ln M$, given by:
\begin{equation}\label{eq:PSfunction}
\frac{d n_h}{d\ln M} = \frac{\rho_m}{M}\nu_c(M)f\left(\nu_c(M)\right) \frac{d\ln \nu_c(M)}{dM}~, 
\end{equation}
where the quantity $\nu_c(M)$, called the peak significance, is defined as
\begin{equation}\label{eq:peak}
\nu_c(M)=\frac{\delta_{cr}}{\sigma(M,z)}=\frac{\delta_{cr}}{D(z)\sigma(M)}~. 
\end{equation}
In eq. \eqref{eq:peak}, $\sigma(M,z)$ denotes the density variance at redshift $z$, evaluated through the linear evolution of $\sigma(M)$ (obtained from eq.~\eqref{eq:variance} to the time of collapse $z$, using the linear growth factor $D(z)$. For the latter we adopt the standard normalization $D(z=0)=D_0=1$. We note again that for the critical overdensity to collapse we use the EDS solution $\delta_{cr}=1.686$~. 

In the original PS theory, the multiplicity function, $f\left(\nu_c(M)\right)$, had the simple form:
\begin{equation}\label{eq:psmult}
\nu_cf\left(\nu_c\right)=\sqrt{\frac{2}{\pi}}\nu_ce^{-\frac{\nu_c^2}{2}}~, 
\end{equation}
which is exact in an EDS universe described by a power-law power spectrum. While this prescription, often referred to as the universal mass function, has been used to describe the halo mass function for a broad range of cosmologies, it lacks the necessary accuracy for predictions in the era of precision cosmology. As a result, ST~\cite{Sheth:1999mn} later introduced an alternative function:
\begin{equation}\label{stmult}
\nu_cf\left(\nu_c\right)=\sqrt{\frac{2}{\pi}}A(p)\left[1+\frac{1}{(q \nu_c^2)^p}\right]\sqrt{q}\nu_ce^{-\frac{q\nu_c^2}{2}}~,
\end{equation}
with $A(p)=\left[1+\pi^{-\frac{1}{2}}2^{-p}\Gamma(0.5-p)\right]^{-1}$ and where $q,p$ are free parameters to be fitted over N-body simulations, reducing to the vanilla PS function for $q=1,p=0$. The best fit pair was initially proposed to be $(q,p)=(0.707,0.3)$ which was later updated to $(q,p)=(0.75,0.3)$, commonly considered to be the ``standard" ST parameters \citep{Sheth:1999mn,Sheth:1999su}. When focusing on higher redshift probes like in our work, however, a slightly different choice of values, $(q,p)=(0.85,0.3)$, has been found to provide a more accurate fit~\cite{Schneider:2020xmf}, which is the one we adopt in order to match the analysis of~\cite{Sabti:2023xwo} that we compare against. We also comment on the fact that we experimented with both choices and found our results to change minimally across the tests, demonstrating the robustness of our analysis against the modeling details of the halo mass function, as was also reported by~\cite{Sabti:2023xwo}. 

To summarize, given a specific choice of the multiplicity function \eqref{eq:psmult}, eqs.~\eqref{eq:radiusdef}-\eqref{eq:peak} provide a prediction for the halo mass function for a given cosmology, that we subsequently use to obtain our galaxy observables of interest from eq. (8) of the main text (MT). Its sensitivity to the underlying cosmological model manifests through the linear matter power spectrum entering eq.~(\ref{eq:variance}), which we obtain from the Boltzmann solver \textsc{CAMB} \cite{Lewis:1999bs} in the $\Lambda$CDM case and additionally using MT-eq. (7) to account for the PBH-induced effects.
        
\section{PBHs from non-Gaussian perturbations and $\mu$ distortion}
\label{app:mu-distortion}

According to the standard hypothesis of PBH formation large-amplitude primordial fluctuations undergo spherical gravitational collapse upon horizon reentry~\cite{Carr:1974nx}. Fluctuations that dissipate via Silk damping~\cite{Silk:1967kq} during the photon diffusion scale, i.e. $5\times 10^4 <z<2\times 10^6$, inject energy in the photon bath and modify the number of photons at different frequencies w.r.t the black-body equilibrium.

Let us start with the probability density function for the primordial curvature perturbations, which can be parameterized as~\cite{Nakama:2016kfq}

\begin{equation}\label{eq:pdf}
P(\zeta)=\frac{1}{2\sqrt{2}\tilde{\sigma} \Gamma\left(1+1/p\right)}\exp \left[-\left(\frac{|\zeta |}{\sqrt{2}\tilde{\sigma}}\right)^p\right]~,
\end{equation} 
where $p=2$ corresponds to the the Gaussian limit of ref.~\cite{Carr:1975qj}.
The variance is  
\begin{equation}\label{eq:sigma}
\sigma^2= \int_{-\infty}^\infty \zeta^2 P(\zeta)d\zeta=\frac{2\Gamma(1+3/p)}{3\Gamma(1+1/p)}\tilde{\sigma}^2~,
\end{equation}
and the abundance of PBH is
\begin{equation}\label{eq:beta}
\beta=\int_{\zeta_c}^\infty P(\zeta)d\zeta =\frac{\Gamma(1/p, 2^{-p/2}(\zeta_c/\tilde{\sigma})^p)}
{2p\Gamma(1+1/p)}~,
\end{equation}
where $\zeta_c\simeq 0.67$ is the threshold for PBH formation~\cite{Harada:2017fjm}, $\Gamma(a)$ is the gamma function and $\Gamma(a,z)$ the incomplete gamma function.

It is shown in ref.~\cite{Chluba:2012we} that a modification to the curvature power spectrum of the form
\begin{equation}\label{eq:delta_PS}
{\Delta \cal P}_\zeta= 2\pi^2 \sigma^2 k^{-2}\delta(k-k_\delta)~,
\end{equation}
which exhibits an extremely sharp feature at some scale $k_\delta$, generates the following $\mu$ distortion
\begin{equation} \label{eq:mu}
\mu\simeq 2.2\sigma^2 \left[
\exp\left(-\frac{\hat{k}_\delta}{5400}\right)
-\exp\left(-\left[\frac{\hat{k}_\delta}{31.6}\right]^2\right)
\right]~.
\end{equation}
where $\hat{k}_\delta=k_\delta \rm Mpc$.

For a PBH population of mass $M_{\rm PBH}$\footnote{Eq.~\eqref{eq:delta_PS} does not necessarily correspond to a monochromatic PBH mass function since critical collapse may broaden the mass spectrum~\cite{Carr:2016hva}.} the scale $k_\delta$ is given by~\cite{Nakama:2016gzw}
\begin{equation}\label{eq:K_delta}
k_\delta\simeq 13 \mathrm{Mpc}^{-1} \gamma^{1/2}\left(\frac{g}{10.75}\right)^{-1/12}\left(\frac{M_{\rm PBH}}{10^{11}M_\odot}\right)^{-1/2}~,
\end{equation}
where $\gamma$ gives the size of the PBH in units of
the horizon mass at formation (we take it to be $\gamma=1$) and $g$ is the number of degrees of freedom of relativistic particles. The initial abundance $\beta$ is related to $f_{\rm PBH}$~\cite{Carr:2009jm,Nakama:2016gzw}, via  

\begin{equation} \label{eq:beta_PBH}
\beta\simeq  6\times 10^{-4}
\left(\frac{g}{10.75}\right)^{1/4}\left(\frac{\Omega_{\mathrm{DM}}}{0.27}\right)^{-1}\left(\frac{M_{\rm PBH}}{10^{11}M_\odot}\right)^{1/2}f_{\rm PBH}~.
\end{equation}
Replacing this expression in eq.~\eqref{eq:beta}, we can solve for $\tilde{\sigma}$ and substitute in eq. \eqref{eq:sigma} to get $\sigma$ and finally calculate $\mu$ using eq. \eqref{eq:mu}.

We notice also that it is not straightforward to connect the expression of eq.~\eqref{eq:pdf} to known models of inflation~\cite{Byrnes:2012yx,Young:2013oia,Franciolini:2018vbk,Atal:2019cdz,Davies:2021loj,Cai:2022erk,Ferrante:2022mui,Pi:2022ysn,Iacconi:2023slv,Huang:2023chx,Hooper:2023nnl}. In more realistic scenarios that exhibit local NGs, one can examine also modifications to the shape of the overdensity and thus the critical threshold for collapse~\cite{Atal:2019cdz, Yoo:2019pma, Kehagias:2019eil, Atal:2019erb}. However, NGs corresponding to values $p<0.5$ cannot be achieved by traditional quadratic $f_{\rm NL}$ and cubic $g_{\rm NL}$ parameters~\cite{Nakama:2017xvq}.

\section{PBH binary distribution during radiation domination}
\label{app:bin_for}
In this appendix, we present the computation of the distribution of binaries as a function of eccentricity and major axis. We follow closely \cite{Sasaki:2018dmp} (see also refs.~\cite{Sasaki:2016jop,Escriva:2022duf} and ref.~\cite{Nakamura:1997sm} for the earlier proposals). We need first to compute the moment during radiation domination when the PBH pair decouples from the expansion of the universe. This occurs when the energy density enclosed by the two PBHs becomes larger than the radiation energy density in the same sphere
\bea 
1+ z_{\rm dec} = (1+ z_{\rm eq}) \bigg(\frac{x_{\rm max}}{x}\bigg)^3~ ,
\eea 
where $x$ is the comoving separation between the two PBHs and $x_{\rm max} = f_{\rm PBH}^{1/3} l_{\rm PBH}(z=0)$. The closest third PBH, at comoving distance $y$, exerts a torque on the binary and prevents head-on collision and transmits angular momentum to the now rotating binary. The distribution of binaries, parameterized by the semi-major axis $a$ and the eccentricity $e$, can be described by 
\begin{align} 
dP \simeq & \, \frac{4\pi x^2 dx}{n_{\rm PBH}^{-1}}\frac{4\pi y^2 dy}{n_{\rm PBH}^{-1}} \Theta( y- x) \Theta (y_{\rm max}- y)~ ,
\nonumber
\\
= & \, \frac{4\pi^2}{3}n_{\rm PBH}^{1/2} f_{\rm PBH}^{3/2}(1+ z_{\rm eq})^{3/2} a^{1/2}e (1-e^2)^{-3/2}de \,da ~ ,
\label{eq:dP}
\end{align}
with $a = \frac{\rho_{c, 0} \Omega_{\rm DM} x^4}{(1+z_{\rm eq})M_{\rm PBH}}$, $e = \sqrt{1- (x/y)^6}$, and $y_{\rm max} = \big(\frac{4\pi}{3} n_{\rm PBH}\big)^{-1/3}$. This leads to a maximum on $a$ and $e$, 
\begin{align}
a_{\rm max} =& \, \frac{x_{\rm max}}{1+z_{\rm eq}} ~ ,
\\
e_{\rm max} = & \, 1- \bigg(\frac{4\pi n_{\rm PBH}}{3}\bigg)^2 \bigg(\frac{(1+z_{\rm eq})M_{\rm PBH}}{\rho_{c, 0} \Omega_{\rm DM}  }a\bigg)^{3/2} ~.
\end{align}

\paragraph*{\textbf{Levels of approximation.}}
The analysis of the binary formation and the derivation of the GW background we presented and used in this letter follows the prescription of ref.~\cite{Sasaki:2016jop}. Firstly, it relies on the three body approximation, and several effects might bring corrections to the rate we estimated. However, it was shown in ref.~\cite{Ioka:1998nz} that those effects are $\mathcal{O}(1)$ corrections and would not substantially alter the results of our analysis. Moreover, for low enough $f_{\rm PBH}\ll 1$, the disruption of binaries formed by close encounter with a third PBH is expected to be small~\cite{Raidal:2018bbj,DeLuca:2020jug}. 

The analyse derived in ref.~\cite{Sasaki:2016jop} which we follow analyzes mergers from PBH binary with masses $M_{\rm PBH}\sim 10^2 M_{\odot}$. Since the relevant physics in the radiation-dominated universe is mass independent, we don't expect any special difference between the properties of PBH binaries with $M_{\rm PBH}\sim 10^2 M_{\odot}$ and the ones in the superheavy regime $M_{\rm PBH}> 10^6 M_{\odot}$, such as for instance the eccentricities and the major axis (see also ref. \cite{Deng:2021gkx,Atal:2020yic}). 

We find that the late-time contribution to binary formation, using the formula in ref.~\cite{Sasaki:2016jop}, is subdominant to the early-time contribution for $M_{\rm PBH} \lesssim 10^9~M_{\odot}$, assuming $f_{\rm PBH} \sim 10^{-2}$ \cite{Sasaki:2016jop}. The latter accounts for the binary formation due to encounter within a dark matter halo, and energy loss through GW emission. At larger masses, $M_{\rm PBH} \gtrsim 10^9~M_{\odot}$, the formula breaks down since there are less than 2 PBHs per dark matter halo, assuming a typical halo mass $M_{\rm halo}\sim 10^{12}~M_{\odot}$. Dedicated studies supported by $N$-body simulations are required for a more precise determination of binary formation at late time in this mass range, considering contributions induced by halo mergers and energy loss through dynamical friction \cite{Begelman:1980vb,Dosopoulou:2016hbg,Ellis:2023owy}. We notice that the latter effect leads to constraints derived in ref.~\cite{Carr:1997cn}, excluding PBHs of mass $M_{\rm PBH} \sim 10^{9}~ M_{\odot}$ unless $f_{\rm PBH} \lesssim 10^{-3}$.

Finally, PBHs can grow in mass due to accretion~\cite{Volonteri:2010wz}. The effect depends on parameters whose values are generically uncertain for all SMBHs, see e.g.~\cite{Outmezguine:2018nce}. We notice that data seems to suggest that accretion becomes inefficient at large masses $M_{\rm PBH} \gtrsim 10^9~\rm M_{\odot}$, e.g. see Fig.~1 in \cite{Hooper:2023nnl}. For those two reasons, we neglect accretion in this analysis.

\begin{figure}[!h]
\centering
 \includegraphics[scale=0.4]{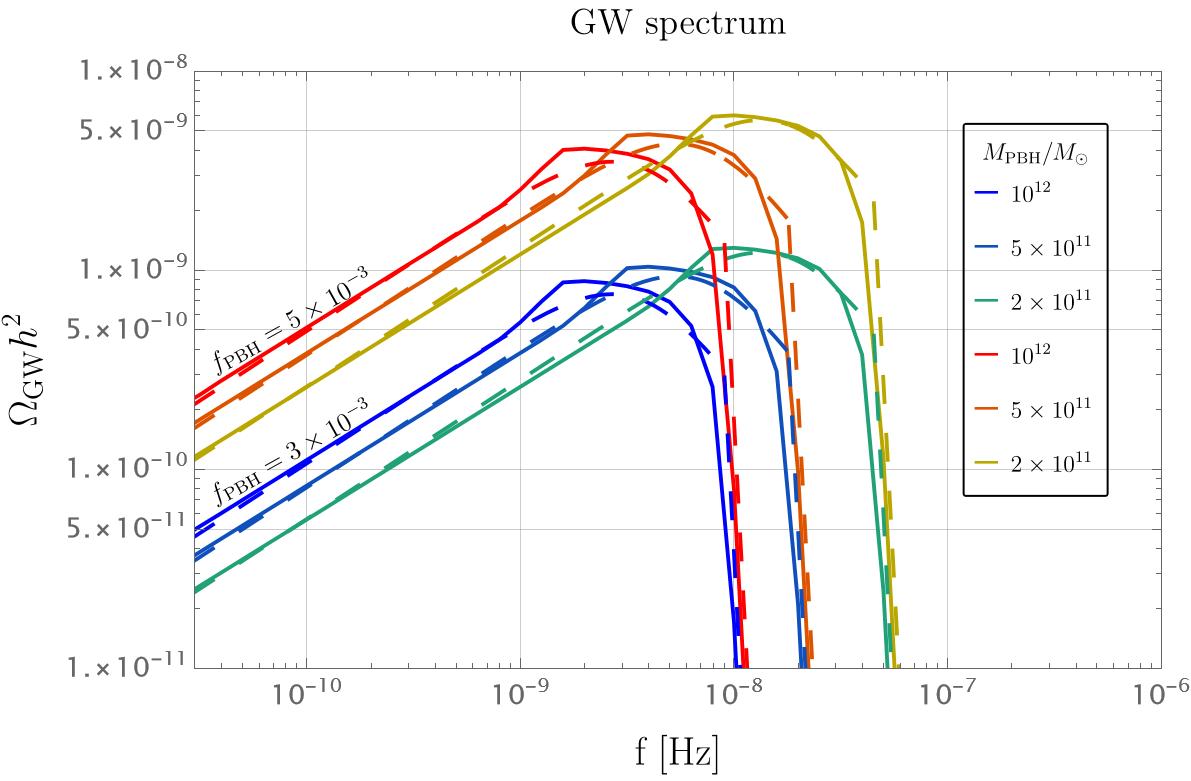}
 \caption{Performance of the fit in eq.~\eqref{eq:fit}. Thick lines designates the full spectrum computed using eq.~\eqref{eq:GW_spectrum_app} and dashed lines designates the fit presented in eq.~\eqref{eq:fit}.}
 \label{fig:fit}
 \end{figure} 
 
\section{Derivation of gravitational wave signal and PTA observations}
\label{app:GW_PBH}

After their formation during radiation era, the two PBHs pertaining to the binary will orbit around each other and lose energy mostly via gravitational radiation. We can estimate their time of merging, assuming $e \to 1$, by~\cite{PhysRev.136.B1224} 
\begin{equation} 
t_{\rm m} (a, e)
\simeq \frac{3}{170}\frac{1}{G^3 M^3_{\rm PBH}} a^4 (1-e^2)^{7/2} ~. 
\end{equation}
This expression of the time of merging $t_{\rm m} (a, e)$ can be inverted to eliminate $a(t_{\rm m} , e)$ in eq.~\eqref{eq:dP} to express the probability distribution as a function of $t_{\rm m}$ and $e$. This leads to the expression \cite{Sasaki:2018dmp}
\begin{equation}
    \frac{dP}{dt_{\rm m}} \simeq \frac{3}{58}  \bigg(\frac{t}{T}\bigg)^{\frac{3}{8}} \frac{1}{t_{\rm m}} \bigg[\frac{1}{(1-e_{\rm up})^{\frac{29}{16}}}-1\bigg] ~,
\end{equation}
where the typical time of merging is 
\begin{equation}
T = \frac{3}{170}\frac{1}{G^3 M^3_{\rm PBH}} \bigg(\frac{3 y_{\rm max}}{4\pi f_{\rm PBH} (1+ z_{\rm eq})}\bigg)^4 ~,
\end{equation}
and the maximal eccentricity reads 
\begin{equation}
e_{\rm up} = \, \begin{cases}
  \sqrt{1- (t_{\rm m}/T)^{6/37}}  \qquad \text{for } t_{\rm m}< t_c ~,
  \\
 \sqrt{1- (4\pi f_{\rm PBH}/3)^2(t_{\rm m}/t_c)^{2/7}}  \qquad \text{for } t_{\rm m}> t_c\, .
\end{cases}
\end{equation} 
The time of the transition between the two regimes is of the form $t_c = T (4\pi f_{\rm PBH}/3)^{37/3}$. 
The final expression for the rate of merging of PBHs can be obtained by multiplying this probability with the PBHs density, $n_{\rm PBH}$,  and we obtain
\begin{align}
\label{eq:merging_app}
\mathcal{R}(t_{\rm m}) \simeq & \, \frac{3 n_{\rm PBH}}{58} \bigg(\frac{t}{T}\bigg)^{\frac{3}{8}} \frac{1}{t} \bigg[\frac{1}{(1-e_{\rm up})^{\frac{29}{16}}}-1\bigg]~.
\end{align}

The gravitational wave amplitude emitted from the merger of two PBHs is then~\cite{PhysRevLett.116.131102}
\begin{equation}
\label{eq:GW_spectrum_app}
h^2 \Omega_{\rm GW} = \frac{f}{\rho_c/h^2 } \int_0^\infty dz \frac{\mathcal{R}(z)}{(1+z)H(z)} \frac{dE_{\rm GW}(f')}{df'} \bigg|_{f'=(1+z)f} \, ,
\end{equation} 
where $dE_{\rm GW}(f')/df'$ is the GW power emitted by an individual binary~\cite{Ajith:2007kx, Ajith:2009bn} 
\begin{align}
\frac{dE_{\rm GW}(f)}{df} = & \, \frac{(G \pi)^{2/3} M_c^{5/3}}{3} 
\nonumber
\\
\times  & \,
\begin{cases}
f^{-1/3}~ , \quad f < f_1 
\\
f^{2/3}f_1^{-1}~ , \quad  f_1 < f < f_2 
\\
f_2^{-4/3}f_1^{-1} \bigg(\frac{f}{1+4(\frac{(f-f_2)}{\sigma})^2}\bigg)^2~ , \quad  f_1 < f < f_2
\end{cases}
\end{align} 
where $M_c$ is the chirp mass, $M_c = M_{\rm PBH}/2^{1/5}$.
We approximate the GW power spectrum in eq.~\eqref{eq:GW_spectrum_app} by the fitting function power broken-law 
\begin{align} 
 \label{eq:fit}
\Omega_{\rm GW} h^2 \simeq    \, \Omega_{\rm peak} S(f) \Theta(2f_{\rm peak} - f)~,
\end{align} 
where the spectral function is
\begin{align} 
 S(f) =   \,\frac{f_{\rm peak}^b f^a }{\bigg(b f^{\frac{a+b}{c}}+ a f_{\rm peak}^{\frac{a+b}{c}}\bigg)^c}~, \nonumber
 \\
a = 0.7, \quad b = 1.5, \quad c = 0.9
 \end{align} 
the peak amplitude is
\begin{align}
 \Omega_{\rm peak}\simeq   \, 0.05 f_{\rm PBH}^3 \bigg(1.5\frac{M_{\rm PBH}}{10^{12} M_{\odot}}\bigg)^{-0.3}~,
\end{align}
and the peak frequency is $f_{\rm peak } \simeq  \, 5000 M_{\odot}/M_{\rm PBH}$. The goodness of the approximation is shown in fig.~\ref{fig:fit}.

\paragraph*{\textbf{Continuous signal.}}
The signal observed at PTAs is a continuous signal. In this paragraph, we explain how to determine if the merger of PBH would procude a continuous signal or a \emph{pop-corn} like signal. The merging rate $\mathcal{R}$ can be related to the comoving number of binaries emitting in the logarithmic interval, in the interval of redshift $[z, z +dz]$, via 
 \bea 
 \frac{dN_{\rm mergers}}{dz d \log f}  = -\mathcal{R}(t) \frac{dV_c}{dz} \frac{d t}{d \log f}~.
 \eea 
 We can further define $\tau_r = dt/d\log {f}$, known as the residence time \cite{Taylor:2021yjx}, as the amount of time that each binary spends emitting in a logarithmic frequency interval, which takes the following form~\cite{Maggiore:2018sht}
 \begin{equation}
 \label{eq:residence_time}
     \tau_r = \frac{5}{96\pi^{8/3}} \left( \frac{1}{GM_c} \right)^{5/3} f^{-8/3} ~ ,
 \end{equation}
 where $M_c = M_{\rm PBH}/2^{1/5}$ in the chirp mass. We can see that $dV_c/dz = 4\pi d^2_c(z)/H(z)$, where $d_c$ is the comoving distance, and thus the number of mergers can be obtained by integrating over $f$ and $z$,
 \bea 
 \label{eq:N_Delta_f}
 N(f,\Delta f) =\hspace{-0.2cm}\ \int^{f_{0}+ \Delta f/2}_{f_0- \Delta f/2}\hspace{-0.1cm}\frac{df}{f}\hspace{-0.2cm}\ \int^{\infty}_0\hspace{-0.2cm}\ dz \frac{ 4\pi d^2_c(z)}{H(z)} R(z)\tau_r  ,~
 \eea 
 where $f_{0} \simeq 2~ \textrm{nHz}$ is the frequency of the first bin and $\Delta f = 14  f_{0}$ the number of bins considered. We notice that the computation depends only very weakly on $\Delta f$. 

 \paragraph*{\textbf{Details of the Bayesian analysis.}}
 In this letter we performed a Bayesian analysis of the PBH mergers interpretation of the PTA data GW signal using eq. \eqref{eq:fit}, which we truncate below the IR cut-off frequency $f_{\rm min}$ when orbital energy is controlled by star slingshot. We impose the condition to have more than one emitting source per frequency bins, $N(\Delta f)>1$ (see eq.~\eqref{eq:N_Delta_f}) as a prior in our analysis. To search for a GW power spectrum from supermassive PBH binaries in the timing residual data, we used the wrapper ${\tt PTArcade}$~\cite{Mitridate:2023oar} which relies on the software tools ${\tt enterprise}$~\cite{enterprise} and ${\tt enterprise\_extensions}$~\cite{enterpriseext} to calculate the likelihood accounting for all the different sources of noise. We included the first 14 frequency bins for NANOGrav 15-year dataset \cite{NANOGrav:2023gor} and the first 13 frequency bins for IPTA DR 2 dataset \cite{Antoniadis:2022pcn}. The chains are generated with the parallel-tempering Markov Chain Monte-Carlo sampler ${\tt PTMCMC}$~\cite{justinellis20171037579}. Finally, we use the software ${\tt GetDist}$~\cite{Lewis:2019xzd} to visualize the posterior distribution, shown with blue and orange ellipses in MT-fig.~1 and fig.~\ref{fig:PBH_exclusion_app}. 

\begin{center}
\begin{figure}
\includegraphics[width=0.9\textwidth]{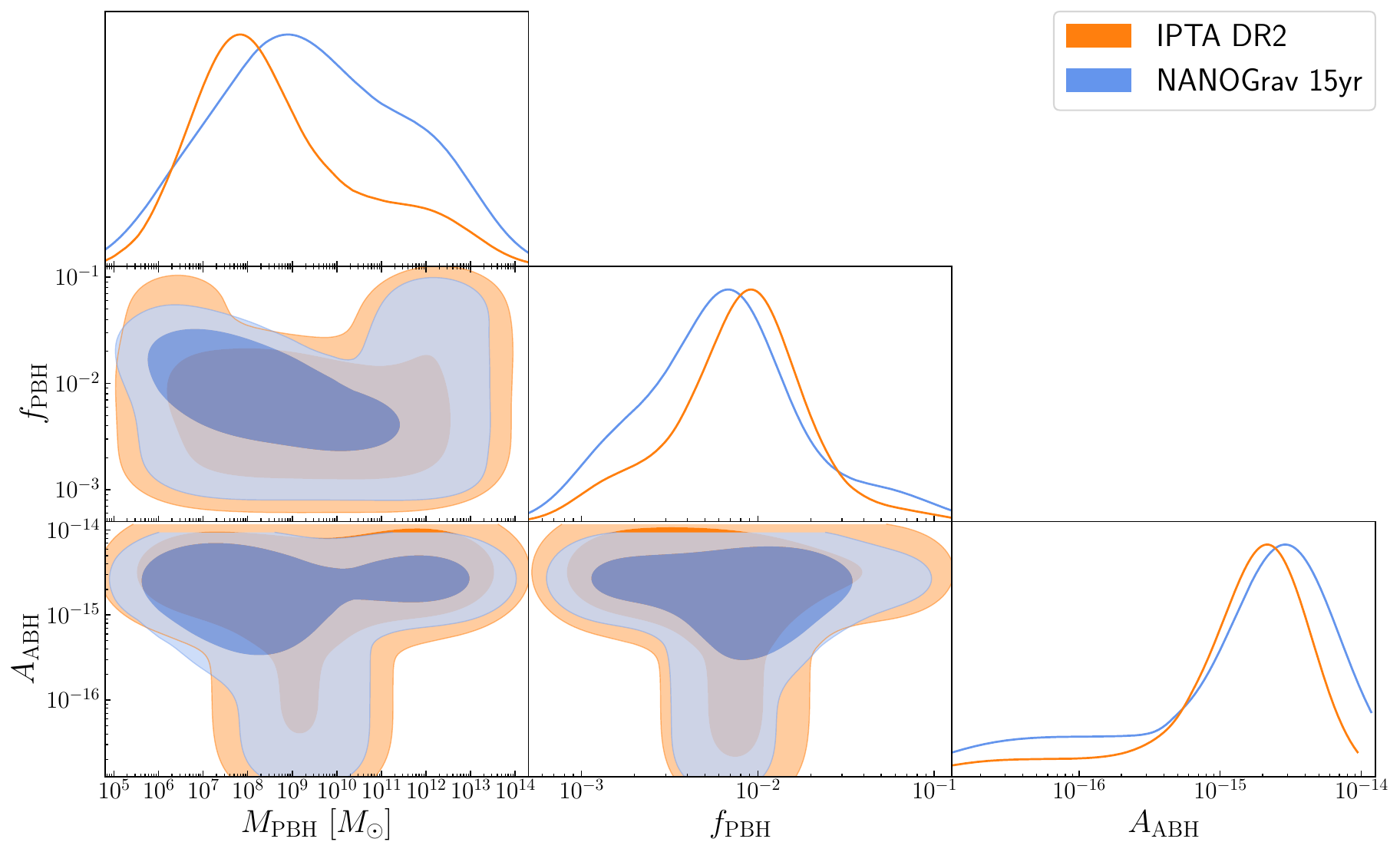}
    \caption{We performed a Bayesian analysis of the scenario where the SGWB observed in PTA is due to a combined population of primordial and astrophysical SMBH, denoted PBH and ABH respectively. We express the $68\%$ and $95\%$ credible interval in terms of the mass $M_{\rm PBH}$ and abundance $f_{\rm PBH}$ of PBHs as well as the strain amplitude $A_{\rm ABH}$ of astrophysical SMBH binary SGWB.\label{fig:SMPBH_vs_SMBH}}
\end{figure}
\end{center}

  \paragraph*{\textbf{Case of SGWB from combined PBHs and ABHs.}}
The favored interpretation of the PTA signal is attributed to SMBH binary of astrophysical origins \cite{NANOGrav:2023hfp,EPTA:2023xxk}. Those are supposed to form at the center of halos in the late universe with a mass correlated to the halo mass \cite{Volonteri:2010wz}.
For completeness, we perform a Bayesian analysis of the combined GW spectrum from primordial and astrophysical black holes (PBHs and ABHs). 
The typical strain spectrum for a set of coalescing, GW-driven SMBH binaries follows a red-tilted power-law \cite{Phinney:2001di}
\begin{equation}
h_c(f) = A_{\text{ABH}} \left(\frac{f}{1~\text{yr}^{-1}}\right)^{-2/3},
\end{equation}
with $A_{\text{ABH}}$ denoting the strain amplitude at a frequency of $1~\text{yr}^{-1} \simeq 3.2\times 10^{-8}~\rm s^{-1}$. When expressed as a fraction of energy density, this relates to a blue-tilted power-law:
\begin{equation}
\Omega_{\text{ABH}}(f) = \frac{2\pi^2}{3H_0^2}f^2 h_c^2(f) \sim f^{2/3}.
\end{equation}
We perform a joint analysis for gravitational waves emanating from both PBH and ABH binary systems. 
 We show the credible intervals in fig.~\ref{fig:SMPBH_vs_SMBH}.
\onecolumngrid\
\twocolumngrid\

\clearpage

\section{PBH parameter space with a higher cut-off}
\label{app:kPBH_cut}

\begin{figure}[h]
\centering
 \includegraphics[scale=0.25]{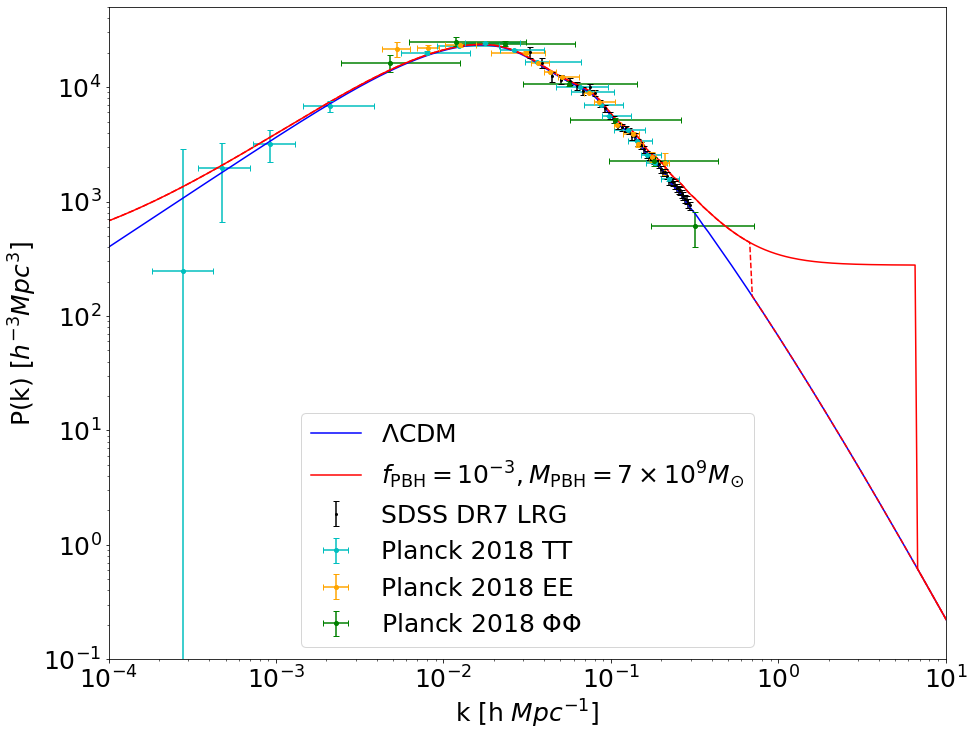}
\caption{The linear matter power spectrum at $z=0$ is plotted for $\Lambda$CDM (blue), for a PBH model with $f_{\rm PBH}=10^{-3}$ and $M_{\rm PBH}=7\times 10^{9} M_{\odot}$ (red dashed) and also using the alternative cut-off $k_{\rm cut} \approx (\bar{n}_{\rm PBH}/f_{\rm PBH})^{1/3}$ (red solid). The points with errorbars correspond to observations of Luminous Red Galaxies (LRG) by the SDSS collaboration \cite{2010MNRAS.404...60R} (black) and the {\it Planck} 2018 \cite{Planck:2018vyg} CMB temperature (cyan), polarization (orange) and lensing (green) power spectra.}
 \label{fig:Pkkcuts}
 \end{figure}

In order to highlight the complementarity between the UV LF and other standard cosmological probes, we plot, in fig.~\ref{fig:Pkkcuts}, the modification to the $\Lambda$CDM power spectrum that is predicted by a supermassive PBH population, together with galaxy observations by the SDSS collaboration \cite{2010MNRAS.404...60R} and the {\it Planck} 2018 \cite{Planck:2018vyg} CMB measurements\footnote{In principle, Lyman-$\alpha$ measurements do access wavemodes in the range $0.1\lesssim k/ \rm{Mpc}^{-1} \lesssim 10$ ~\cite{Chabanier:2019eai, Murgia:2019duy, Blas:2021mqw}. However, it is unclear how to extrapolate the Lyman-$\alpha$ bound into the region $M_{\rm PBH} > 10^5 M_{\odot}$~\cite{Carr:2020erq,Carr:2020gox}. As a result, we do not include them in Fig. 3.}. Even though the observational landscape is soon expected to significantly change with the advent of Stage-IV surveys such as DESI~\cite{2013arXiv1308.0847L}, Euclid~\cite{Laureijs:2011gra} and the Rubin Observatory LSST~\cite{Abate:2012za}, we note that such deviations are currently not yet ruled out by standard constraints, highlighting the importance of the UV LF for this study.

\begin{figure}[h]
\centering
 \includegraphics[scale=0.5]{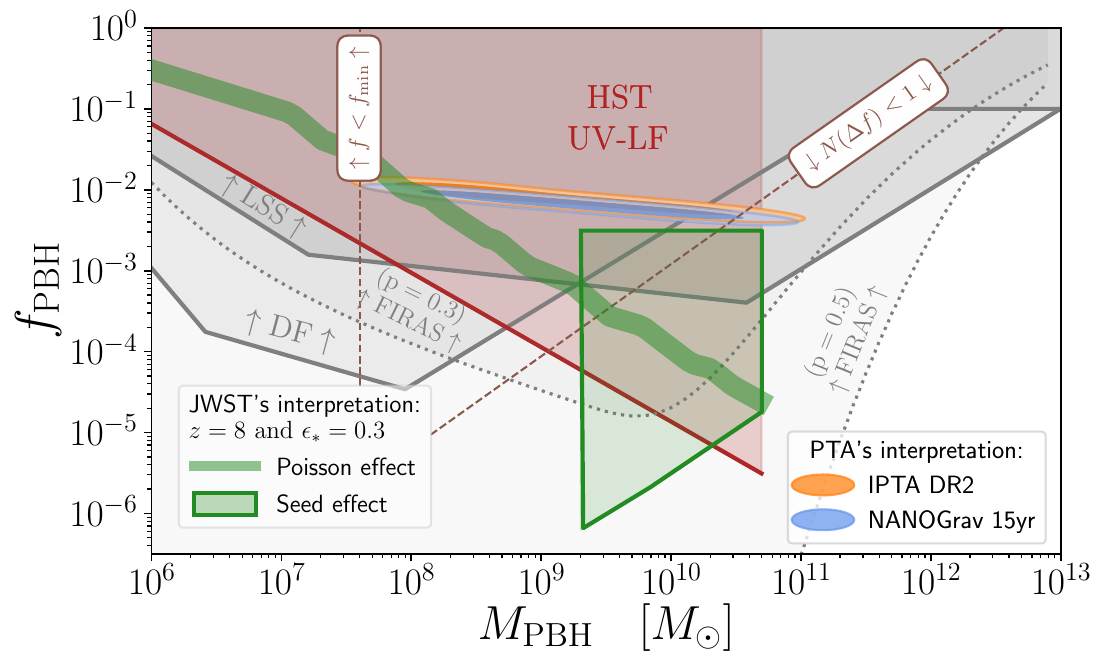}
 \caption{Same as fig.~1 in the main text with a different wavenumber cut-off $k_{\rm cut} \approx (\bar{n}_{\rm PBH}/f_{\rm PBH})^{1/3}$ as assumed in ref.~\cite{Liu:2022bvr}. }
 \label{fig:PBH_exclusion_app}
\end{figure}

We then proceed to discuss the results of the main analysis using the same cut-off $k_{\rm cut} \approx (\bar{n}_{\rm PBH}/f_{\rm PBH})^{1/3}$ on the isocurvature component in MT-eq. (7) as it is assumed in the heuristic approach of ref.~\cite{Liu:2022bvr}. To illustrate the difference from the cut-off $k_{\rm cut} = \bar{k}_{\rm PBH}$ used in the MT, we plot in fig.~\ref{fig:Pkkcuts} the modification to the power spectrum predicted by both cut-offs. As it is evident, this choice allows for higher wavenumbers to contribute and thus the boost to the halo mass function is significantly increased. As a result, the curve in MT-fig.~1 where the JWST observations are satisfied as well as the bound derived from the UV LF are both displaced towards lower $f_{\rm PBH}$ values, as clearly seen in fig.~\ref{fig:PBH_exclusion_app}. 

It is noteworthy that in this case the solution based on the seed effect is partially subjected to the UV LF bound. The underlying assumption is that at intermediate scales between the linear- and non-linear regimes the two effects co-exist and even though the seed effect dominates, the Poisson may still contribute at a smaller percentage (see also fig.~6  in ref.~\cite{Inman:2019wvr}, but notice that the authors employ linear perturbation theory only above $k_{\rm cut} = \bar{k}_{\rm PBH}$).

\FloatBarrier

\bibliographystyle{JHEP}
{\small
\bibliographystyle{apsrev4-2}
}
\end{document}